\documentclass[12pt]{article}

  \let\LARGE=\large
  \let\Large=\normalsize
  \let\large=\normalsize

  \newcommand{\beq}{\begin{equation}}
  \newcommand{\eeq}{\end{equation}}
  \newcommand{\beql}[1]{\begin{equation}\label{eq:#1}}

  \newcommand{\beqa}{\begin{eqnarray}}
  \newcommand{\eeqa}{\end{eqnarray}}
  \newcommand{\beqas}{\begin{eqnarray*}}
  \newcommand{\eeqas}{\end{eqnarray*}}

  \newcommand{\bC}{{\bf C}}

  \newcommand{\bS}{{\bf S}}

  \newcommand{\cH}{{\cal H}}

  \newcommand{\cM}{{\cal M}}
  
  \newcommand{\cO}{{\cal O}}

  \newcommand{\al}{\alpha}

  \newcommand{\mb}{\mbox}

  \newcommand{\rh}{\rho}

  \newcommand{\De}{\Delta}

  \newcommand{\Eq}[1]{Eq.~(\ref{eq:#1})}
  \newcommand{\Ga}{\Gamma}

  \newcommand{\Tr}{\mbox{\rm Tr}}

  \newcommand{\beqan}{\begin{eqnarray*}}
  \newcommand{\beqar}[1]{\begin{equation}\label{#1}\begin{array}{l}}

  \newcommand{\eeqar}{\end{array}\end{equation}}

\newcommand{\bra}[1]{\langle#1|}
\newcommand{\ket}[1]{|#1\rangle}

\newcommand{\og}{\overline{g}}
\renewcommand{\Ga}{G}
\begin{document}
\begin{titlepage}
\begin{flushright}
TU- 704\\
quant-ph/0401187
\end{flushright}
\ \\
\begin{center}
\LARGE
{\bf
Quantum Estimation by  Local Observables 
}
\end{center}
\ \\
\begin{center}
\Large{
Masahiro Hotta${}^\ast$ and Masanao Ozawa${}^\dagger$    }\\
{\it
${}^\ast$
Department of Physics, Faculty of Science, Tohoku University,\\
Sendai, 980-8578,Japan\\
hotta@tuhep.phys.tohoku.ac.jp \\

\ \\

${}^\dagger$
Graduate School of Information Sciences, Tohoku University,\\
Sendai, 980-8579, Japan \\
ozawa@math.is.tohoku.ac.jp
}
\end{center}

\begin{abstract}
Quantum estimation theory provides optimal observations for various 
estimation problems for unknown parameters in the state of the system
under investigation.
However, the theory has been developed under the assumption 
that every  observable is available for experimenters.
Here, we generalize the theory to problems in which 
the experimenter can use only locally accessible observables.
For such problems, we establish a Cram{\'e}r-Rao type inequality
by  obtaining an explicit form of the Fisher information as 
a reciprocal lower bound for the mean square errors of
estimations by locally accessible observables.
 Furthermore, we explore various local quantum 
estimation problems for composite systems, 
where non-trivial combinatorics is needed for
obtaining the Fisher information.
\end{abstract}

\end{titlepage}

\section{Introduction}
\ \\

In many experimental situations,
we are not allowed to have a large number of 
 data enough to determine unknown parameters such as coupling constants of
hypothetical interactions.
 In some cases, the number may be  fairly small
and it is crucial to theoretically explore the best estimator for the
parameter from the small number of our available data.
 The problem becomes prominent for quantum systems, since optimal 
estimation must be well reconciled with inevitable quantum uncertainty
arisen from available observables and unknown parameters. 
 In such situations, 
 the quantum estimation theory can  play a significant role;  
for  detailed reviews, we refer to 
 Helstrom \cite{Hel76} and Holevo\cite{Holevo}. 
The theory provides the best observation on the system 
 for the estimation with the minimum value of the estimate error.

Although the ordinary quantum estimation theory is certainly  powerful  
 for many estimation problems, the theory  
 includes an implicit assumption, which is not 
 realistic in some of the practical experiments. 
The assumption is that 
 every observable of the system is available for
  the observer or the experimenter.
 Contrary to the assumption  practically available
observables are often  restricted. 
 For instance, it is a common situation in experiments that
a particle is contained inside the laboratory 
at the origin of the time and that the experimenter can only use measuring
devices inside the laboratory.
However, according to the time evolution the particle
may go out of the laboratory, so that 
the ability of estimating the state parameter  is restricted to
measuring devices inside the laboratory for the later time. 
 Another example is found in elementary particle  physics. 
 It usually  happens due to
 the limit of  the present technology of measurement 
 that our apparatus can probe only low-energy portions of the total
Hilbert space with visible signals. 
Thus the observables are certainly restricted. 
  In such situations,  the observable provided by the ordinary theory
for the best estimate may in general not be available . 
Then, the question becomes relevant as to
 what is the best estimate among those which are accessible  
{\em only by use of restricted observables}. 
 Let us generally call such estimations 
 local quantum estimations. 
 
In this paper, we elaborate the formulation of quantum estimation
theory for local quantum estimation problems on an unknown parameter $g$.  
For the restricted density operators measured  by our apparatus,
 a Fisher information is introduced.
 Then, we prove the quantum Cram{\'e}r-Rao type inequality
 for the local quantum estimation for $g$. 
 The observable is specified that attains the equality and 
  yields the best local quantum estimate for $g$ by its measurement. 
It is also pointed out that there exist 
nontrivial aspects in the analysis of the local quantum estimation
 for the composite system of identical subsystems.
 In that case we have two natural estimations 
 and corresponding two Fisher informations
for the unknown parameter $g$. 
The first alternative 
 takes a simple form to apply, but may give a smaller value of the Fisher information. 
The second alternative is able to give a larger value of 
 the Fisher information and generates a better estimate for $g$, 
 but have a pretty complicated form to deal with, 
 compared to the first alternative.  Especially, 
 calculation of  the second  Fisher information 
 requires solving independently evolutions of many descendant operators.

In section 2, a brief review on the standard quantum estimation theory
 is given. 
 In section 3, we discuss more physically the quantum estimation problem,
  including the biased-estimator case. 
 Several expected advantages of the quantum estimation are also reviewed. 
In  section 4,  we introduce the notion of local quantum estimation problems. 
 In  section 5, a quantum Cram{\'e}r-Rao inequality for 
 local quantum estimations is established.  
In  section 6, Fisher information is discussed for unnormalized pure states. 
 In  section 7, we reveal nontrivial aspects of local quantum estimations 
for composite systems. 
In  section 8, two general formulations  are proposed for 
 local quantum estimations for the composite system. 
 In  section 9, a formula which is useful for the 
 evaluation of one of two sorts of Fisher information 
 for composite systems is given. 
 In  section 10, we apply our formulation for local quantum estimations 
  to a decaying two-level system with a small unknown parameter. 
  In the final section, we summarize  our results of this paper.

\section{Cram\'{e}r-Rao bound for quantum estimators}
\ \\

Let us briefly review quantum estimation theory
  in this section. For a detailed review we refer the reader to 
  Helstrom \cite{Hel76} and Holevo\cite{Holevo}. 
  
Let $\bS$ be a closed quantum system described by
a Hilbert space $\cH$.
We assume that the Hamiltonian $H_{tot}$
has a constant real parameter $g\in\Ga$, 
i.e., $H_{tot}=H_{tot}(g)$, 
where $\Ga$ is the set of possible values of $g$.
The evolution equation for the density operator 
$\rho_{tot}$ of $\bS$  is given by 
\begin{equation}
i\hbar \partial_{t}\rho_{tot} = [H_{tot},\ \rho_{tot}].
\end{equation}
Then, the density operator of $\bS$ at a given time $t$
depends on the time $t$ and the parameter $g$, i.e., 
\beq\label{eq:tot}
\rh_{tot}=\rh_{tot}(g,t).
\eeq

We shall consider the following quantum estimation 
problem for the parameter $g$.
In order to estimate the parameter $g$ in $H_{tot}(g)$, 
we assume that one measures an observable $A$ 
at time $t$,
and the output $\overline{g}$ is taken as the estimate 
for $g$.
Thus, the observable $A$ plays the role of the estimator
of this statistical estimation problem. 

By the Born statistical formula,  the expectation value 
of the measurement output $\og$ in the state 
$\rh_{tot}(g)=\rh_{tot}(g,t)$ is given by
\beqa
E_{g}[A]=\Tr[A\rh_{tot}(g)].\label{1e}
\eeqa
Then,  $E_{g}[A]$ is the mean of one's estimate $\og$ for
the given true value $g (\in\Ga)$. 
The variance of the estimate $\og$ 
for the true value $g (\in\Ga)$ is given by
\beqa
V_{g}[A]=(\De_{g} A)^{2}=\Tr[A^{2}\rh_{tot}(g)]
-\Tr[A\rh_{tot}(g)]^{2},
\eeqa
where $\De_{g} A$ is the uncertainty of observable
$A$ in the state $\rh_{tot}(g)$. 
The mean-square error $\epsilon^2_g [A]$ of the estimate $\og$ 
for the true value $g$ is defined by
\begin{equation}
\epsilon^2_g [A] := E_g [(A-g)^2].
\end{equation}
By a simple manipulation, we obtain the relation 
\begin{equation}
\epsilon^2_g [A] =V_g [A] +(E_g [A] -g)^2. \label{mse}.
\end{equation}
 The estimator $A$ is called {\em unbiased}, if the mean
estimate is correct for any possible values $g (\in\Ga)$,  
i.e.,
\begin{equation}
E_{g}[A]=g \label{ub}
\end{equation}
for any $g (\in\Ga)$. 
In this section, we shall confine our attention to 
unbiased estimators $A$.  
As seen in eqn(\ref{mse}), for the unbiased $A$,  
the variance $V_{g}[A]$ represents 
the mean-square error of the estimate $\og$ for the
true value $g (\in\Ga)$:
\begin{equation}
\epsilon^2_g [A]=V_g [A].
\end{equation}

The lower bound for $V_{g}[A]$ is given by the 
well-known quantum Cram\'{e}r-Rao inequality as follows.
The symmetric logarithmic derivative (SLD) $L(g)$
for  $\rh_{tot}(g)$ is defined as a self-adjoint operator
satisfying
\beqa\label{eq:SLD}
\partial_{g}\rh_{tot}(g)=
\frac{1}{2}[\rh_{tot}(g)L(g)+L(g)\rh_{tot}(g)].
\eeqa
Note here that 
\begin{equation}
E_g [L(g)]= \Tr[\rh_{tot} (g) L(g)] =0,\label{2e}
\end{equation}
due to the normalization condition of $\rh_{tot}$.
By the above relation, the SLD may not be determined 
uniquely; however, any two solutions $L_{1}(g)$
and  $L_{2}(g)$ satisfy the the relation \cite{FN95}
\beqa
L_{1}(g)\rh_{tot}(g)=L_{2}(g)\rh_{tot}(g).
\eeqa
Thus, the operator $L(g)\rh_{tot}(g)$ is uniquely 
determined.
The Fisher information $J_{g}$ of the parameter 
$g$ in $\rh_{tot}(g)$ is uniquely defined by
\beqa 
J_{g}=\Tr[L^{2}(g)\rh_{tot}(g)]. \label{fi}
\eeqa
Then, every unbiased estimator $A$ satisfies  
the quantum Cram\'{e}r-Rao inequality \cite{Hel76,Holevo}
\beqa
\epsilon^2_g [A] =V_{g}[A]\ge\frac{1}{J_{g}}.\label{crie}
\eeqa
A simple proof is given in the appendix.

\section{More Physical Review of Quantum Estimation}
\ \\

In this section, we shall discuss real experimental procedures for estimating
the unknown parameter $g$ in the framework of quantum estimation
theory given above.
In real experiments, rigorous unbiased estimators that satisfy 
 eqn(\ref{ub}) globally in the parameter space of $g$ are usually not 
available for experimenters from technical reasons. 
 Instead, only biased estimators are available.
Even in such real situations, as we shall show in the following,
the quantum estimation theory described in the previous section
plays an active role. 
We also discuss some advantages of quantum estimation theory
to provide more efficient methods in several estimation problems
in physics.

We shall first consider a quantum estimation 
 for a parameter $g$  in $H_{tot}(g)$ performed by a  measurement of 
 a general observable $A$ for a single sample at time $t$. Here $A$ is not 
 assumed unbiased.  
 Assume that we get an outcome $\bar{a}$ in the measurement
and we make the estimate $\bar{g}$ for $g$ as a function of the outcome $\bar{a}$.
Usually, this function $\bar{g}=\phi(\bar{a})$ is obtained by the following 
way.
The relation
\begin{equation}
E_{g}[A]=a \label{e5}
\end{equation}
between the true value $g$ and the mean output $a$ from the measurement
can be often solved theoretically as a function
\begin{equation}\label{phi}
g=\phi(a).
\end{equation}
Usually, sensible experiments are designed to possess a suitable domain 
 ${\cal G } (\subseteq G)$, which includes interesting values of $g$, and
 to allow the relation
\begin{equation}
\bar{g}=\phi(\bar{a}),
\end{equation}
applying the above function $\phi$ to the measurement output $\bar{a}$,
gives a good estimate $\bar{g}$ for the given true value $g (\in {\cal G})$
from the output of  the single $A$  measurement. 
 Note that, due to the quantum nature, 
the observable $A$ generally possesses 
 nonzero values of the variance $V_g [A]$.
  The variance has a close relationship to the problem;
 to what extent the estimate $\bar{g}$ can be trusted. 
 For example, if $V_g[A]$ is extremely small, 
 then, even in the single measurement, 
the observed value  $\bar{a}$ must be almost equal
  to the expectation value $E_g [A]$ for the correct value of $g$. 
  Consequently, the estimate $\bar{g}$ has to almost coincide with 
 the correct $g$ value. 
From the viewpoint of the dimensional analysis,
 it is rather straightforward to introduce an expected error $\delta g_g [A]$ 
of the estimate $\og_{\bar{a}} [A]$
for the true value of $g$ as follows. 
\begin{eqnarray}
(\delta g_g [A])^2 :=\frac{V_g [A]}{(\partial_g E_g [A])^2}.
\label{er1}
\end{eqnarray}

Now, we shall give a justification of the above estimate and 
error evaluation from the viewpoint of quantum estimation theory. 
When $V_g [A]$ is small enough, 
 a domain ${\cal G}$ in the space $G$ of possible values
 can be chosen so narrow  
 that $E_g [A]$ is linearly expanded around a physically interesting 
 value $g_o (\in {\cal G})$:
\begin{equation}
E_g [A] = E_{g_o} [A]+\partial_g E_{g}[A]|_{g=g_o}
 (g-g_o) +O\left((g-g_o)^2 \right).
\end{equation} 
Here it is quite useful to remind 
that a single measurement of the observable $A$ 
simultaneously implies  a single measurement of an observable $f(A)$, 
where $f(x)$ is an arbitrary real function of $x$. 
 If an output $\bar{a}$ of the observable $A$ is obtained, 
 it is interpreted that 
 an output $f(\bar{a})$ is 
 observed for the observable $f(A)$ in the same measurement. 
 In what follows, in order to make a useful choice of $f$,
let us impose on the function $f$ 
 the locally unbiased condition:
 \begin{equation}
E_g [f(A)] = g +O\left((g-g_o)^2 \right). \label{luc}
\end{equation}
 Then it is noticed  that eqn(\ref{luc}) is satisfied 
 for $g\in {\cal G}$ by a linearized function $f(x)$
such that
\begin{equation}\label{f(x)}
f(x) = \frac{x}{\partial_g E_g [f(A)] |_{g=g_o}} 
+g_o -\frac{E_{g_o} [f(A)]}{\partial_g E_g [f(A)] |_{g=g_o}}. 
\end{equation}
By linearity of $f$, for the mean output $a=E_{g}[A]$  we have 
\begin{equation}
f(a)=f(E_{g}[A])=E_g [f(A)],
\end{equation}
so that from eqn(\ref{phi}) we have
\begin{equation}
\phi(\bar{a})=f(\bar{a}) +O\left((g-g_o)^2 \right).
\end{equation}
Thus,  $\bar{g}=\phi(\bar{a})$ is now reproduced  
  by substituting the output $\bar{a}$ into the function $f(x)$.
By use of eqn(\ref{mse}) and eqn(\ref{luc}), 
the mean-square error of the estimate $\bar{g}$ as
the output $f(\bar{a})$ of  
the $f(A)$ measurement  is evaluated as
\begin{equation}
\epsilon^2_g [f(A)]=V_g [f(A)] +O\left( (g-g_o)^4 \right). \label{mv}
\end{equation}
From eqn(\ref{f(x)}), the variance of $f(A)$ is  evaluated as
\begin{equation}
V_g [f(A)] =\frac{V_g [A]}{(\partial_g E_g |_{g=g_o} [A])^2}
=(\delta g_g [A])^2 
\left(1
+O\left( g-g_o \right)
\right),\label{ve}
\end{equation}
thus, it is verified due to eqn(\ref{mv}) and eqn(\ref{ve}) that
\begin{equation}
\epsilon^2_g [f(A)] =(\delta g_g [A])^2
\left(1 +O\left( g-g_o \right)\right).
\end{equation}
 Hence the validity of the error evaluation by $(\delta g_g [A])^2$
definition has been shown.

It is a significant result from quantum estimation theory 
 that even for the biased observable $A$, 
the quantum Cram\'{e}r-Rao inequality can be proven: 
\begin{equation}
(\delta g_g [A] )^2 \geq \frac{1}{J_g}.\label{e111}
\end{equation}
 Here the Fisher information is defined by eqn(\ref{fi}) and 
 the equality can be achieved by taking $A\propto L(g)$ for each value of $g$. 
 This can be shown by adopting not $A$ but 
 the local unbiased operator $f(A)$, and 
 returning to the general argument for the unbiased case 
 in  section 2. It is also possible to 
 prove by using the biased observable $A$ straightforwardly.  
 The proof can be seen in the appendix.

Next we shall discuss physically relevant cases, the N-samples systems,
  by naturally extending the single-sample argument. 
Let us take a  
 composite system 
 which consist of N identical {\bS} subsystems.
Assume here 
that the density operator
 is independent and identically distributed; 
\begin{equation}
\rho_{tot}^{(N)} (g,t)= \rho(g,t)^{\otimes N}.
\end{equation}
Now the average estimators $\bar{A}^{(N)}$ defined by
\begin{equation}
\bar{A}^{(N)}
:= \frac{1}{N}\sum^N {\bf 1}\otimes \cdots \otimes A \otimes {\bf 1}
\cdots \otimes {\bf 1}
\end{equation}
 are available. 
 According to the quantum law of large numbers \cite{LargeN},
 the measurement data for $\bar{A}^{(N)}$ 
 are going to be normally distributed
  with the average $E_g[A]$ and the standard deviation 
  $\left(\frac{V_g[A]}{N}\right)^{1/2}$ when the number $N$ becomes large.
Since the expected error is solely
 a pull back of the quantum deviation of
  the observable, we can trust the estimate, 
\begin{equation}
g =
\bar{g} \pm
 \delta g_{\bar{g}} [\bar{A}^{(N)}],
\end{equation}
 in the 1-$\sigma$ precision for the large number cases.

Now let us discuss the Fisher information for the N-samples cases.
 The SLD for the composite system $L^{(N)}$ are
  defined for general density
  operators $\rho_{tot}^{(N)}$ by 
\begin{equation}
\partial_{g}\rh_{tot}^{(N)}=
\frac{1}{2}[\rh_{tot}^{(N)}L^{(N)}+L^{(N)}\rh_{tot}^{(N)}].
\end{equation}
For  independent and identically distributed (i.i.d.) 
 density operators, 
 it is easily derived that the composite SLD $L^{(N)}$ 
 is given by
\begin{equation}
L^{(N)} =\sum^N 
{\bf 1}\otimes \cdot \otimes{\bf 1} \otimes L\otimes {\bf 1} \otimes
\cdots \otimes {\bf 1}, \label{e12}
\end{equation}
where ${\bf 1}$ is the identity operation and $L$ is the SLD for the 
subsystem defined by eqn(\ref{eq:SLD}). This result yields the following  
simple relation for the Fisher information of the composite system
 $J^{(N)}_g =\Tr [(L^{(N) })^2 \rho^{(N)}]$:
\begin{equation}
J^{(N)}_g =NJ_g^{(1)}.
\end{equation}

By virtue of  the Cram\'er-Rao inequality, it is easily noticed that 
the optimized estimation for $g$ in a single measurement of the composite
system is achieved by adopting the average estimator $\bar{L}^{(N)}
=\frac{1}{N} L^{(N)}$.
 The expected error is given by
\begin{equation}
\delta g_{g} [\bar{L}^{(N)}] =\frac{1}{\sqrt{N J_g^{(1)}}}.
\end{equation}
This coincides with  the usual error  of the estimation 
for $g$  based upon N independent data of  measurements 
of $L^{(1)}$ for the N  subsystems. 
However, stress  that there is no need  to 
 measure N times the estimator $L^{(1)}$ for each subsystem $\bS$ 
 to achieve the estimate. In the quantum estimation, just 
 one measurement of the single observable $\bar{L}^{(N)}$ 
 yields the best estimate.  Other
 relative-difference components like $L\otimes{\bf 1} \cdots \otimes{\bf 1}
 -{\bf 1}\otimes L\cdots\otimes{\bf 1}$ 
 remain unmeasured. This  saves effectively  
 the number of processes in the estimation and exposes 
  an advantage of the quantum 
 estimation.

When the entanglement between the subsystems is available,
 it is possible  \cite{Hayashi} 
 that the large $N$ behavior of $\delta g_g [\bar{L}^{(N)}]$ can be
 improved  beyond the $1/\sqrt{N}$ factor as
\begin{equation}
\delta g_g [\bar{L}^{(N)}] \propto \frac{1}{N}.
\end{equation}
This reveals another advantage of the quantum estimation.

\section{Quantum Estimation by Local Observables} 
\ \\

We shall now consider the following constraints
on the quantum estimation problem discussed above.
In the above general formulation, we have assumed that
every observable $A$ of the system $\bS$
is available for our measurement to fix the $g$ value.  
However, in practice the available
observables are restricted.  
For instance, it is a common situation in experiments that
a particle described as the system $\bS$ is contained inside the laboratory 
at the origin of the time and that we can only use measuring
devices inside our laboratory.
However, according to the time evolution the particle
may go out of the experimental apparatus or our laboratory, 
 so that for the general $t$,
our ability of estimating the parameter $g$ is restricted
by the measuring devices inside the laboratory.

Let $\cM$ be a subspace of $\cH$.
The projection of $\cH$ onto $\cM$ is denoted by $P$.
In this paper, we consider the following two constraints.

(i) The initial state is supposed to be supported by $P$,
i.e., 
\beqa\label{eq:local condition}
\rho_{tot}(t=0) =P\rho_{tot}(t=0) P.
\eeqa

(ii) The available observables ${\cal O}$ for our measurements 
 are restricted to those of the form
\beqa
{\cal O}=PXP+y(I-P),
\eeqa
where $X$ is an arbitrary observable on $\cH$ and $y$
is an arbitrary real number.

Let  $\{\ket{a}\}$ be the orthonormal basis of 
$\cM$ and $\{\ket{\al}\}$ be the orthonormal basis
of $\cH$ extending  $\{\ket{a}\}$, i.e,
$\{\ket{a}\}\subseteq\{\ket{\al}\}$.
Then, condition \Eq{local condition} is equivalent to
the relation
\begin{equation}
\rho_{tot}(t=0) 
=\sum_{a,a'} |a\rangle \langle a |\rho_{tot}(t=0)|a' \rangle
\langle a'|.
\end{equation}
Thus, the density operator initially has only  
matrix elements inside $\cM$, and 
according to the time evolution, the density operator
$\rho_{tot}$ may have matrix elements outside 
of $\cM$.

In the case of estimating the parameter $g$ by
observing a particle $\bS$ initially localized in
a box using the measuring devices effective 
only inside the box, the subspace $\cM$ corresponds to the
space of wave functions localized in the box.
In this case, the assumption that the particle is initially localized
inside the box is represented by condition (i).
Since we assume that we know that the particle
inside the box at the origin of the time, 
by measuring later, for instance, the weight of the box , 
we can measure $I-P$ via a negative result of the measurement.
Since the measuring devices are only effective inside
the box, the measuring interaction couples only with 
the observable of the form $PXP$ so that it is natural
to assume that they can measure only observables of
the form $PXP$.
Therefore, the set of available observables are
considered to be restricted to those given by
condition (ii).

Initially the density operator $\rho_{tot}$ have only 
matrix elements in $\{\ket{a}\}$.  However, in the course of
the time evolution, $\rho_{tot}$ can have matrix elements outside
of $\{\ket{a}\}$.  For any nonnegative time $t\geq 0$, we define 
the accessible density operator $\rho_\parallel(t)$ 
for the subspace $\cM$ by
\beqa
\rho_\parallel(t)=P\rho_{tot}(t)P.\label{e2}
\eeqa
Obviously, $\rho_\parallel$ has the matrix representation 
\begin{equation}
\rho_\parallel(t)=\left[ \langle a |\rho_{tot}(t) | b \rangle \right].
\end{equation}
Then, by the corresponding properties of $\rho_{tot}(t)$, the 
operator  $\rho_\parallel(t)$ is positive and 
satisfies 
\begin{equation}
0\leq \Tr \rho_\parallel(t) \leq 1.
\end{equation}
In what follows, we shall consider the time domain of $t$ from
$t=0$ to the time just before $t=t_*$ such that 
$\Tr\rho_\parallel (t=t_*) =0$, where we allows the case 
$t_* =\infty$,
so that we have
\begin{equation}
0<\Tr \rho_\parallel(t) \le 1,
\end{equation}
for $t\in [0,t_*)$.

From condition (ii), the available estimators $A$ on $\cM$ are naturally 
restricted and satisfy the relation
\begin{equation}
A=A_{\parallel} +a_{\perp} ({\bf 1}-P)=A^\dagger\label{40},
\end{equation}
where
\begin{equation}
A_\parallel =PA_{\parallel}P.\label{e3}
\end{equation}
Using the definitions in eqns (\ref{e2}), (\ref{40}) and (\ref{e3}),
it can be shown that
 the expectation value of the available estimator $A$ is given by
\begin{equation}
\langle A \rangle =\Tr(\rho_{tot}(t) A) =\Tr(\rho_\parallel(t)A_{\parallel})
+a_{\perp}(1-\Tr\rho_\parallel(t) ).\label{e1}
\end{equation}

In order to define rigorously 
 the notion of the ``local'' estimators $\tilde{A}$ 
corresponding to  the available estimators $A $ in the restricted situation, 
let us extend the space $\cM$ to a one-dimension-larger 
 Hilbert space $\tilde{\cM}$ by adding to the basis of $\cM$ 
 a normal vector $|B\rangle$ 
orthogonal to every $|a\rangle$,i.e., 
$\tilde{\cM}:=\cM\oplus {\bf C}|B\rangle$. 
  Then, the local estimators $\tilde{A}$ acting 
 on $\tilde{\cM}$, which corresponds to the available estimator in eqn(\ref{40}), are defined by
\begin{equation}
\tilde{A}=A_{\parallel}
+a_{\perp} |B\rangle \langle B |. \label{70}
\end{equation}
In particular, note that 
\beqa
\widetilde{1-P}= |B\rangle \langle B |.\label{e4}
\eeqa
Since the state $|B\rangle$ 
represents the  inaccessible states  by our local observation as seen 
in eqn(\ref{e4}), we call $|B\rangle$  the blank state.
 
Further let us introduce 
the local density operator $\rh$ acting on $\tilde{\cM}$ 
and corresponding to $ \rho_\parallel (=P\rho_{tot}P) $
 by
\begin{eqnarray}
\rho=\rho_\parallel
+(1-\Tr\rho_\parallel ) |B\rangle \langle B |.\label{e20}
\end{eqnarray}
It is easily seen that $\rho$ is positive and of unit trace.
By a simple manipulation,  
the expectation value of the available estimator $A$ in eqn(\ref{e1}) 
 can be  
 reexpressed by use of the local estimator $\tilde{A}$ 
 and the local density operator $\rho$ 
  as 
\begin{equation}
\langle A \rangle =\Tr(\rho \tilde{A} ).
\end{equation}

\section{Cram\'{e}r-Rao Bound for Local Quantum Estimators}
\ \\

In what follows, we shall consider the quantum
Cram\'er-Rao inequality for the
quantum estimation problem for the coupling constant $g$
in the Hamiltonian $H_{tot}(g)$ by using {\em only} local measuring devices.
By the time evolution, the local density operator $\rho=\rho(t,g)$ introduced
 in the previous section
depends on the time $t$ and the parameter $g$.
Now we assume that one measures a local estimator $\tilde{A}$ at time
$t$, and the output $\bar{a}$ determines  the estimate $\bar{g}_{\bar{a}} [A]$ 
 for $g$ 
via the relation:
\beqas
E_{\bar{g}_{\bar{a}} [A]}[A]=\Tr[\tilde{A}
\rho(t,\bar{g}_{\bar{a}} [A]) ]=\bar{a}.
\eeqas
Stress that, in this estimating process of $g$, 
 we are allowed to use only local estimators $\tilde{A}$ 
instead of arbitrary observables in the theory.

The variance of the local observable $\tilde{A}$ for the correct $g$
 value 
is certainly given by
\beqa
V_{g}[\tilde{A}]=\Tr
[\tilde{A}^{2}\rho(t,g)]-\left(\Tr[\tilde{A}\rho(t,g)]\right)^{2}.
\eeqa
Then it is required at a given time $t$ 
  to find the minimum value of the expected error defined by 
\begin{equation}
\delta g_g [\tilde{A}] :
=\sqrt{\frac{V_g [\tilde{A}]}{(\partial_g E_g[\tilde{A}])^2}}.
\end{equation}
It is shown that this problem is resolved by use of  a solution of  
 the problem on the 
estimate for the parameter $g$  
 by arbitrary observables $\tilde{{\cal O}}$ on 
$\tilde{\cM}$ as follows.

We define the local SLD $\tilde{L}(g)$ on $\tilde{\cM}$ 
 for an arbitrary local density operator $\rho(g)$ in eqn(\ref{e20}) 
as a self-adjoint operator satisfying
\begin{equation}
\partial_g \rho(g) =\frac{1}{2}(\tilde{L}(g) \rho +\rho
 \tilde{L}(g)),\label{e16}
\end{equation}
\begin{equation}
\tilde{L}(g)^\dagger =\tilde{L}(g).\label{e18}
\end{equation}
Since $\Tr\rho(g)=1$ for any $g\in G$, we have 
\begin{equation}
\Tr[\tilde{L}(g)\rho(g) ]=0.
\end{equation}
It is easy to construct a solution of eqn(\ref{e16}) by introducing 
 a SLD operator $L(g)$ on $\cM$ for the accessible density operator $\rho_\parallel$. The SLD $L(g)$ on $\cM$ is
 defined by 
\begin{equation}
\partial_g \rho_\parallel 
=\frac{1}{2}(L(g) \rho_\parallel +\rho_\parallel
 L(g)),\label{e17}
\end{equation}
\begin{eqnarray}
&&
L(g)^\dagger =L(g),
\\
&&
PL(g)P =L(g).
\end{eqnarray}
Due to the fact that $P\rho_\parallel P =\rho_\parallel $,
 we can find, at least, a solution of eqn(\ref{e17}) 
 for the SLD with $PL(g)P =L(g)$. 
Once the SLD $L(g)$ is given, then it is proven  by a simple algebra that 
the operator defined by 
\begin{equation}
\tilde{L}(g)=L(g)+ \partial_g 
\ln [1-Tr\rho_{\parallel}(g)]|B\rangle \langle B| \label{e19}
\end{equation}
 satisfies eqns(\ref{e16}) and (\ref{e18}), thus it is a SLD 
 on $\tilde{\cM}$ for $\rho(g)$.

Here it is carefully  noted that we may have
\begin{equation}
\Tr[L(g)\rho_{\parallel}(g) ]  \neq 0,
\end{equation}
since the trace of $\rho_{\parallel}(g)$ is not necessarily normalized.

The operator $\tilde{L}(g)$ is determined uniquely up to the support of 
$\rh(g)$; any two solutions $\tilde{L}_{1}(g), \tilde{L}_{2}(g)$
satisfy $\tilde{L}_{1}(g)\rh(g)=\tilde{L}_{2}(g)\rh(g)$.
The Fisher information $J_{g}$ of the parameter $g$ in 
$\rho(g)$ is uniquely defined by
\begin{equation}
J_{g}=\Tr[\tilde{L}(g)^2 \rho(g)]
=\Tr[L(g)^2 \rho_\parallel(g)] +
\frac{\left[\Tr[ L(g)\rho_\parallel(g)]\right]^2}{1-\Tr[\rho_\parallel(g)]},
\label{34}
\end{equation}
where we have used eqns(\ref{e20}) and (\ref{e19}). 
Then, for the 
 arbitrary observables $\tilde{\cal O}$ on $\tilde{\cM}$
 we have the quantum Cram\'er-Rao inequality
\begin{equation}
(\delta g_g [\tilde{{\cal O}}] )^2 
:=\frac{V_{g}[\tilde{{\cal O}}]}{(\partial_g E_g [\tilde{{\cal O}}])^2}
\geq \frac{1}{J_{g}},
\end{equation}
where $J_g$ is given by eqn(\ref{34}).
In order to apply the result to our local estimator problem,
 it is crucial to
 notice that the SLD 
 in eqn(\ref{e19}) takes the precise form of the local estimator  
 on $\tilde{\cM}$ in eqn(\ref{70}). 
 Therefore, the equality can be attained by a local estimator.
This indicates that 
  the following quantum Cram\'er-Rao inequality for arbitrary
 local estimators $\tilde{A}$ on $\tilde{\cM}$ really holds
  for the local density operators $\rho$ corresponding 
  to the accessible density operators 
  $\rho_\parallel (g) (=P\rho_{tot}(g) P)$:
\begin{equation}
(\delta g_g [\tilde{A}] )^2 
=\frac{V_{g}[\tilde{A}]}{(\partial_g E_g [\tilde{A}])^2}
\geq \frac{1}{J_{g}},\label{cr}
\end{equation}
where the Fisher information $J_g$ is given by eqn(\ref{34}). 
For a given $g\in G$, the equality is attained by a local estimator  
$\tilde{A}_o (g)$ such that 
\begin{equation}
\tilde{A}_o (g)\propto \tilde{L}(g)=L(g)
-\frac{Tr[L(g)\rho_\parallel(g)]}{1-Tr\rho_\parallel(g)}|B\rangle\langle B|.
\end{equation}
Note that the local estimators which give the minimum expected error such as $\tilde{A}_o$ 
 are  unique only up to a factor and an additive term proportional to the identity operator. For instance, an estimator such that
\begin{equation}
\tilde{A}_o '(g) \propto L(g) +
\frac{Tr[L(g)\rho_\parallel(g)]}{1-Tr\rho_\parallel(g)}P,
\end{equation}
which has no matrix element for the blank state, also attains
the equality.

\section{The Fisher Information for Unnormalized Pure States }
\ \\

In physics, it often happens that the measurement device 
 is able to probe only a small part 
  of the physical  states of the total system.
 Even in such situations, non-unitary formulations are sometimes  
  available.  The state vectors $|\Psi(t)\rangle$ 
 are governed by equations of motion with non-Hermitian 
 Hamiltonians and 
 evolve deterministically in the  subspace $\cM$, which is accessible 
 by the experimental devices. Such examples are found 
 in the various fields of physics, 
 including 
 the scattering problems with weak absorption of quanta 
  in the nuclear physics and the quantum optics,
  the flavor-oscillation studies 
   in the elementary particle physics and so on. 
  The information about  
 the coupling constant $g$ in the equations of motion is 
 imprinted on the state vectors $|\Psi(t,g)\rangle$ 
  during the time evolution.
 
Let us evaluate the Fisher information for the pure state 
$|\Psi (t,g)\rangle$.
The accessible density operator for the pure state reads 
\begin{equation}
\rho_\parallel 
(t,g) =|\Psi(t,g) \rangle \langle \Psi (t,g) |,\label{36}
\end{equation}
where $\Tr[\rho_\parallel (0,g)] =1 $
 and at an advanced time $t(>0)$ the following relation holds:
\begin{equation}
0 < \Tr[\rho_\parallel (t,g)] \leq 1.
\end{equation}
We define the SLD operator $L$ on $\cM$, in the same way discussed 
in the previous section, 
 for the accessible density operator $\rho_\parallel$. Note that 
 the operator $L$ is not uniquely determined due to the 
 purity of $\rho_\parallel$, however, the ambiguity is not relevant at all for the Fisher information, as commented in the previous section.
 It is shown that we have  a SLD, 
\begin{equation}
L=\frac{2}{\Tr[\rho_\parallel]}
\partial_g \rho_\parallel 
-\frac{\Tr[\partial_g \rho_\parallel]}
{(\Tr[\rho_\parallel] )^2}\rho_\parallel,
\end{equation}
as a simple representative and 
 the Fisher information itself is uniquely evaluated by
\begin{equation}
J=4\left(
\langle \partial_g \Psi |\partial_g \Psi \rangle
-
\frac{\left|Im \langle \Psi |\partial_g \Psi \rangle \right|^2}
{\langle \Psi |\Psi \rangle}
\right)
+4\frac{\left|Re \langle \Psi |\partial_g \Psi \rangle\right|^2}
{1-\langle \Psi |\Psi \rangle},\label{unps}
\end{equation}
where $|\partial_g \Psi\rangle :=\partial_g |\Psi (t,g)\rangle$.
This result is an extension of that in the reference \cite{FN95}, 
 where the normalized pure state theory is analyzed.
The relation enables us to evaluate easily the Fisher information
 for many unnormalized  pure state theories.

 In eqn (\ref{unps}), one may worry about the apparent divergence
  of the third term at $\langle \Psi |\Psi \rangle=1$, because 
  the state evolves initially from the normalized state.
 However, for ordinary physical systems, the early behavior of the
 norm $\langle \Psi |\Psi \rangle$ is given by
\begin{equation}
\langle \Psi (t,g)|\Psi (t,g)\rangle \sim 1-\alpha(g) t^2,
\end{equation}
where $\alpha$ is a positive function of $g$. 
Thus the third term is evaluated in the early era as
\begin{equation}
4\frac{\left|Re \langle \Psi |\partial_g \Psi \rangle\right|^2}
{1-\langle \Psi |\Psi \rangle}
\sim \frac{(\partial_g \alpha (g))^2}{\alpha(g)} t^2.
\end{equation}
Hence, the limit $t\rightarrow +0$ of eqn(\ref{unps}) 
exists without any problems.

\section{Problems of the Composite System}
\ \\

In the local estimation problem, some nontrivial aspects appear in the
 composite system analysis. 
  Suppose a system $\bS$. 
 Let us assume our measuring device for  $\bS$ 
 is able to access only a subspace
 $\cM$ of the Hilbert space of $\bS$. 
 Later  let $P$ denote the projection operator onto $\cM$,
  and$D_{\cM}$ denote the dimension of the subspace $\cM$.  
 The accessible density operator on $\cM$ 
 is denoted by $\rho_\parallel$.  
 The  operator $\rho_\parallel (t,g)$ evolves in the subspace $\cM$ 
 and becomes dependent on the coupling constant $g$ in the equation of motion.
 Let us consider a composite system
 $\bS^{\otimes N}$ composed of N identical $\bS$ subsystems. 
 For instance, 
 suppose that an independent and identically distributed (i.i.d.) 
 initial condition
 is set for the total density operator $\rho_{tot}(0)$ 
 of the composite system. Also assume that the unitary evolution of the total system is factorized, i.e., $U^{(N)}(t) =U(t)^{\otimes N}$. 
 Even in such a simple situation, it can be pointed out that 
 we have, at least, two natural 
  alternatives for the estimation of $g$ as follows. 
  
  The first alternative is rather simple. 
  In the procedure, one  firstly calculates the accessible density operator
  $\rho^{(N)}_{\parallel }=P^{\otimes N} \rho_{tot}P^{\otimes N}$ 
  for $\bS^{\otimes N}$  which is reduced to a direct product defined by 
\begin{equation}
\rho^{(N)}_{\parallel } (t,g) :=\rho_\parallel^{\otimes N} (t,g).
\end{equation}
A local density operator for the accessible density operator 
 $\rho^{(N)}_{\parallel }$
  can be defined straightforwardly by
\begin{equation}
\rho_1^{(N)} := \rho^{(N)}_{\parallel } 
+(1-\Tr\rho^{(N)}_{\parallel } ) |B\rangle \langle B |,
\end{equation}
where $|B\rangle$ is the blank state.  
 Let  $j^{(N)}$ denote the Fisher information 
 based upon the first local density operator $\rho_1^{(N)}$.

The  estimation problem in the composite systems is nontrivial 
 because we may have a construction of another local density
  operator for $\bS^{\otimes N}$. 
We are able to define at first the local density operator for
 each subsystem $\bS$. For the i-th subsystem $\bS_i$, 
 the local density operator $\rho_i$ corresponding
  to $\rho_{\parallel i}$ is written as
\begin{eqnarray}
\rho_i =\rho_{\parallel i}
+(1-\Tr\rho_{\parallel i} ) |B_i \rangle \langle B_i |,
\end{eqnarray}
where $|B_i \rangle $ is the blank vector for the i-th subsystem $\bS_i$.
 Then we can define naturally 
 the second local density operator $\rho^{(N)}_2$ for the
  composite system $\bS^{\otimes N}$ by a direct product as follows:
\begin{equation}
\rho^{(N)}_2 :=\prod^N_{i=1}\otimes \rho_i .
\end{equation}
Let $J^{(N)}$  denote the Fisher information  based upon 
 $\rho^{(N)}_2$. By construction, the Fisher information $J^{(N)}$ 
 for the i.i.d. density operator is calculated as
\begin{equation}
J^{(N)} =NJ^{(1)}.
\end{equation}

As seen above, there exist two independent Fisher informations for the composite system.  Then, 
it is an important question; 
 which alternative  of the formulation 
  gives us a more precise estimate for $g$, that is, which Fisher
 information, $j^{(N)}$ or $J^{(N)}$, is larger than another.  
 The problem should be addressed 
for the general initial conditions for the density operators, 
beyond the above i.i.d. situations. 
Note first that
the operators $\rho_1^{(N)}$ act on the Hilbert space ${\cM}^{\otimes N}
\oplus {\bC}|B\rangle$ and the dimension of ${\cM}^{\otimes N}
\oplus {\bC}|B\rangle$ is given by
 $D_1 =(D_{\cM})^N +1$.
 On the other hand,
 the operators $\rho_2^{(N)}$ act on the Hilbert space 
 $\tilde{{\cM}}^{\otimes N}$ 
and the dimension of $\tilde{\cM}^{\otimes N}$ is given by
 $D_2 =(D_{\cM} +1)^N$. Since $D_2 > D_1$ always holds, it is naively
 expected that the second Fisher information $J^{(N)}$ 
  is not less than the first Fisher information $j^{(N)}$. 
  This guess can be
   proved affirmatively  by use of the monotonicity 
  argument for the Fisher information as will be mentioned later.

The above argument has been limited to the i.i.d.cases.
In order to analyze the  composite-system estimation generally, we must
 extend the above two formulations. Especially, nontrivial
  analyses are required to define  $J^{(N)}$.
   These are formulated in  section 8.

In  section 9, it is also pointed out  
  that evaluation of the larger Fisher information $J^{(N)}$ 
  requires solving time evolutions of  
  various density operators corresponding to different initial conditions.  
   Such a feature
    does not appear in the evaluation of both the standard Fisher 
information in the usual  cases 
 and the smaller Fisher information $j^{(N)}$ 
 in the local estimation.

\section{General Formulation for the Composite System}
\ \\

The available estimators for the composite system 
 $\bS^{\otimes N}$ now reads 
\begin{equation}
A^{(N)} =
\sum_{k_1\cdots k_N} 
\omega_{k_1 \cdots k_N} A_{k_1} \otimes\cdots\otimes A_{k_N},\label{80}
\end{equation}
which is just a natural extension of eqn(\ref{40}). Here $A_{k}$
 denote the available estimators for the subsystem $\bS$, 
  which take the form in eqn
 (\ref{40}) and $\omega_{k_1 \cdots k_N}$ are real coefficients. 
The corresponding extension of eqn(\ref{70}) is also possible.
 The local estimator $\tilde{A}^{(N)}$ corresponding to 
 the available estimator $A^{(N)}$ is defined by
\begin{equation}
\tilde{A}^{(N)} =
\sum_{k_1\cdots k_N} 
\omega_{k_1 \cdots k_N} \tilde{A}_{k_1} \otimes\cdots\otimes \tilde{A}_{k_N},
\end{equation}
where $\tilde{A}_k$ are the local estimators corresponding to $\tilde{A}$ 
in eqn(\ref{70}).

In order to define the two Fisher informations $j^{(N)}$ and $J^{(N)}$
 beyond the i.i.d. condition,  let us consider
  the most general local density  operator 
 $\rho_{tot}^{(N)} (0) = P^{\otimes N} \rho_{tot}^{(N)} (0) P^{\otimes N}$
 as the initial total density operator. 
In the unitary time evolution of the total system, 
\begin{equation}
\rho_{tot}^{(N)} 
(t,g) =U^{(N)} (t,g) \rho_{tot}^{(N)} (0) U^{(N)\dagger} (t,g),
\end{equation}
 the total density operator becomes to have
 matrix elements between the inaccessible states. 
 
 Even for the general initial conditions,
  the definition of the first Fisher information $j^{(N)}$ 
 is essentially unchanged. 
  Let us introduce the accessible operators $\rho_{\parallel}^{(N)}$
  by reducing the total density operator $\rho_{tot}^{(N)}$ as
\begin{equation}
\rho_{\parallel}^{(N)} =P^{\otimes N} \rho_{tot}^{(N)} P^{\otimes N}.
\end{equation}
For the accessible density operator $\rho^{(N)}_\parallel$,
 a SLD operator $L^{(N)}$ is defined by
\begin{equation}
\partial_g \rho_\parallel^{(N)} 
=\frac{1}{2}\left[
L^{(N)}  \rho^{(N)}_\parallel +\rho^{(N)}_\parallel
 L^{(N)} 
 \right],
\end{equation}
\begin{eqnarray}
&&
(L^{(N)})^{\dagger} =L^{(N)} ,
\\
&&
P^{\otimes N} L^{(N)}P^{\otimes N} =L^{(N)}.
\end{eqnarray}
According to eqn(\ref{34}), 
 the Fisher information $j^{(N)}$ is defined straightforwardly 
as follows.
\begin{equation}
j^{(N)}
:=\Tr\left[\left(L^{(N)}\right)^2 \rho_\parallel^{(N)}\right] +
\frac{\left(\Tr\left[ L^{(N)} 
\rho_\parallel^{(N)}\right]\right)^2}{1-\Tr\left[\rho_\parallel^{(N)}\right]}.
\end{equation}

 Next, in order to define the second Fisher information $J^{(N)}$,
  what we want is a proper definition of a  
  local density operator
 $\rho^{(N)}$ acting on $\tilde{{\cM}}^{\otimes N}$ such that 
 the total density operator $\rho_{tot}^{(N)}$ 
 is reduced into $\rho^{(N)}$.
 Here, it is quite natural to impose that expectation 
  values of all the available observables $A^{(N)}$ for $\rho_{tot}^{(N)}$ 
  are equivalent to
   those  of the corresponding local observables $\tilde{A}^{(N)}$ 
   for $\rho^{(N)}$:
\begin{equation}
\Tr[ A^{(N)} \rho_{tot}] 
=\Tr[ \tilde{A}^{(N)} \rho^{(N)} ].\label{ee2}
\end{equation}
 By some manipulations it is soon noticed that  the above constraint is really 
 satisfied by defining the local density operator $\rho^{(N)}$  as follows. 
Let index $\alpha_{j}$ for $j=1,\ldots,N$ below
 take index $a_{j}$ for states in $\cM$
 or the index $B$ for the blank state.
 Then the matrix elements of $\rho^{(N)}$ on  
 $\tilde{{\cM}}^{\otimes N}$ are given by
\begin{eqnarray}
&&
\langle \alpha_1 \alpha_2 \cdots \alpha_N |
\rho^{(N)}|\alpha_1 ' \alpha_2 '\cdots \alpha_N '\rangle
\nonumber\\
&=&
\prod_{j=1}^N
\left[
\delta_{\alpha_j B} \delta_{\alpha_j ' B}
+(1-\delta_{\alpha_j B} )(1-\delta_{\alpha_j ' B} )
\right]
\nonumber\\
&&\times
\sum_{x_1 \cdots x_N}
\prod^N_{k=1}
\left[ 
\delta_{\alpha_k B}\delta_{x_k 1}
+
(1-\delta_{\alpha_k B} )\delta_{x_k 0}
\right]
\prod^N_{k'=1}
\left[ 
\delta_{\alpha_{k'} ' B}\delta_{x_{k'} 1}
+
(1-\delta_{\alpha_{k'} ' B} )\delta_{x_{k'} 0}
\right]
\nonumber\\
&&\times
\Tr[(
P_{1,x_1} \otimes P_{2,x_2} \otimes \cdots \otimes P_{N,x_N})
\rho_{tot}], \label{91}
\end{eqnarray}
where $\Tr$ stands for the trace operation on $\cH^{\otimes n}$.
For $m=1,\ldots,N$, the subscript $x_m$ 
takes $0$ or $1$ and the operator $P_{m,x_{m}}$
is defined by 
\begin{eqnarray}
&&
P_{m,x_{m}} =\ket{a'_{m}}\bra{a_{m}},\quad\mb{if $x_{m}=0$}
\\
&&
P_{m,x_{m}}={\bf 1}-P,\quad\mb{if $x_{m}=1$}. 
\end{eqnarray}
By construction the Hermicity of the operator $\rho^{(N)}$ is trivial. 
Further, taking $A^{(N)} ={\bf 1}^{\otimes N}$ in eqn(\ref{ee2}) yields 
the normalization condition:
\begin{equation}
\Tr[\rho^{(N)}] =1.\label{ee1}
\end{equation}
The positivity of $\rho^{(N)}$ is also proven as follows.
Suppose an arbitrary vector $|\Psi\rangle$ 
on $\tilde{{\cM}}^{\otimes N}$:
\begin{eqnarray}
|\Psi\rangle &=&
\sum_{\alpha_1\cdots \alpha_N} 
C_{\alpha_1 \cdots \alpha_N} 
|\alpha_1 \cdots \alpha_N \rangle
\nonumber\\
&=&
\sum_{(i_1 \cdots i_k)} |\Psi_{[i_1 \cdots i_k]} \rangle,
\end{eqnarray}
where 
\begin{equation}
|\Psi_{[\o]}\rangle =
\sum_{a_1\cdots a_N} 
C_{a_1 \cdots a_N}
| a_1 \cdots a_N \rangle,
\end{equation}
\begin{equation}
|\Psi_{[1]}\rangle
=
\sum_{a_2\cdots a_N} \
C_{Ba_2 \cdots a_N}
| B a_2 \cdots a_N \rangle 
\end{equation}
and so on.
 Then, using the definition of $\rho^{(N)}$ 
  in eqn (\ref{91}), 
 the expectation values of $\rho^{(N)}$ for the arbitrary
 state vectors $|\Psi\rangle$ are evaluated as follows.
\begin{eqnarray}
\langle \Psi |\rho^{(N)}|\Psi \rangle
&=&
\sum_{i_1 \cdots i_k} \langle \Psi_{[i_1 \cdots i_k]}|
\rho^{(N)} |\Psi_{[i_1 \cdots i_k] }\rangle
\nonumber\\
&=&
\sum_{i_1 \cdots i_k} \Tr[
\tilde{P}_{[i_1 \cdots i_k]}
\rho^{(N)} ]
\nonumber\\
&=&
\sum_{i_1 \cdots i_k} \Tr[
P_{[i_1 \cdots i_k] }
\rho_{tot} ] ,
\end{eqnarray}
where $\tilde{P}_{[i_1 \cdots i_k]}=
|\Psi_{[i_1 \cdots i_k]} \rangle\langle\Psi_{[i_1 \cdots i_k]}|$ and
 $P_{[i_1 \cdots i_k]}$ are defined by replacing $|B\rangle\langle B|$'s 
  in the operator $\tilde{P}_{[i_1 \cdots i_k]}$ to ${\bf 1} -P$.
 Noting that the operators $P_{[i_1 \cdots i_k]}$
  can be expressed as $P_{[i_1 \cdots i_k]} 
 =\sum_\beta 
 |\Phi_{\beta,[i_1 \cdots i_k]} \rangle\langle\Phi_{\beta,[i_1 \cdots i_k]}|$
 by use of vectors $|\Phi_{\beta,[i_1 \cdots i_k]} \rangle$ 
 in the total Hilbert
  space, it is proven that
\begin{eqnarray}
\langle \Psi |\rho^{(N)}|\Psi \rangle
=
\sum_{\beta, i_1, \cdots i_k} \langle \Phi_{\beta,[i_1 \cdots i_k]}|
\rho_{tot}^{(N)} 
|\Phi_{\beta,[i_1 \cdots i_k]} \rangle
 \geq 0.
\end{eqnarray}
Taking account of the normalization condition in eqn(\ref{ee1}),
 this implies the  positivity of the operator $\rho^{(N)}$.

Since $\Tr [\rho^{(N)}] =1$, 
we can define in the usual way 
 a SLD operator ${\cal L}$ for the local density operator
 $\rho^{(N)}$:
\begin{equation}
\partial_g \rho^{(N)}
=\frac{1}{2} ({\cal L}
\rho^{(N)}
 +\rho^{(N)}
 {\cal L} ).
\end{equation}
Then, the Fisher information $J^{(N)}$ is defined by
\begin{equation}
J^{(N)}=\Tr\left[\rho^{(N)} {\cal L}^2 \right].
\end{equation}

Now let us  comment on the inequality $J^{(N)} \geq j^{(N)}$,
 using the monotonicity of the Fisher information.
  The point is that there exists a mapping $R$ 
  of the density operators defined in $\tilde{\cM}^{\otimes N}$ onto
  the density operators defined in  ${\cM}^{\otimes N}\oplus {\bC}|B\rangle$.
Let us denote $P_\parallel$
 a projection operator onto the subspace of vectors 
 in $\tilde{\cM}^{\otimes N}$ that 
  do not include the blank states at all. Denote $P_\perp$ 
  a projection operator onto the subspace of vectors that include more than one
  sub-blanck vectors $|B_i\rangle$. 
It should be noted that 
\begin{equation}
P_\parallel
\rho^{(N)}
P_\parallel
=
\rho^{(N)}_\parallel.
\end{equation}
Let us define the mapping $R$ as follows.
\begin{eqnarray}
R[\rho^{(N)} ]&=&
P_\parallel \rho^{(N)} P_\parallel 
+\Tr\left[ P_\perp \rho^{(N)}\right] 
|B\rangle\langle B|
\nonumber\\
&=&\rho^{(N)}_{\parallel} 
+\Tr\left[ P_\perp \rho^{(N)}\right]
|B\rangle\langle B|.
\end{eqnarray}
By definition, it is clear that the mapping is linear and 
of unit trace:
\begin{equation}
\Tr\left[ R[\rho^{(N)} ] \right]
=\Tr[P_\parallel \rho^{(N)}] +\Tr[P_\perp \rho^{(N)}] =1.
\end{equation}
It is also easily seen that this mapping is completely positive,
since so are $\rho^{(N)}\mapsto P_\parallel \rho^{(N)} P_\parallel $
and $\rho^{(N)}\mapsto\Tr\left[ P_\perp \rho^{(N)}\right]$.

Using  the relation $\Tr[\rho^{(N)}] =1$, we obtain 
\begin{eqnarray}
\varrho &:=&
R[\rho^{(N)} ]
\nonumber\\
&=&
\rho^{(N)}_\parallel 
+(1-\Tr[\rho^{(N)}_\parallel ])
|B\rangle\langle B|.
\end{eqnarray}
  Then, the first Fisher information $j^{(N)}$ is given by
$\Tr[ \varrho^{(N)} (\tilde{L})^2 ]$, where
 $\tilde{L}$ is the SLD operator corresponding to $\varrho^{(N)}$. 
  According to the monotonicity theorem for the Fisher information 
  \cite{Monotone},
 it must be satisfied under the projective mapping $R$ that
  $J^{(N)} \geq j^{(N)}$.
This result does not depend on whether the total density operators 
$\rho^{(N)}_{tot}$ are factorized or entangled.

It is worth noting that the information $J^{(N)}$ possesses a decomposition
 representation. 
Let us consider an arbitrary subsequence $(i_1 ,i_2 ,\cdots ,i_n)$ of
 the sequence $(1,2,3,\cdots,N)$.
Define that
$\rho_{[i_1 ,i_2 ,\cdots ,i_n]}$ 
 is a $(dim {\cal M})^{N-n}\times (dim {\cal M} )^{N-n}$ matrix which is composed of components of $\rho^{(N)}$ 
  with $\alpha_{i_m} =\alpha_{i_m}'=B$ for $m=1,\cdots, n$. 
  The followings are examples.
\begin{equation}
\langle a_2 a_3 \cdots a_N |
\rho_{[1]}
|a_2 ' a_3 ' \cdots a_N ' \rangle
:=
\langle B a_2 a_3 \cdots a_N |
\rho^{(N)}
|B a_2 ' a_3 ' \cdots a_N ' \rangle, \label{90}
\end{equation}
\begin{equation}
\langle a_2 a_4 \cdots a_N |
\rho_{[1,3]}
| a_2 ' a_4 '\cdots a_N ' \rangle
:=
\langle B a_2 B a_4 \cdots a_N |
\rho^{(N)}
|B  a_2 ' B a_4 ' \cdots a_N ' \rangle,
\end{equation}
\begin{equation}
\langle a_1 |
\rho_{[2,3,\cdots,N]}
|a_1 ' \rangle
:=
\langle a_1 B B \cdots B |
\rho^{(N)}
|a_1 ' B B \cdots B \rangle.
\end{equation}
Note that the empty subsequence $\o$ corresponds to the accessible
  density operator: 
\begin{eqnarray}
\langle a_1 a_2 \cdots a_N |
\rho_{[\o]}
|a_1 ' a_2 ' \cdots a_N ' \rangle
=
\langle a_1 a_2 \cdots a_N |
\rho_\parallel^{(N)} 
|a_1 ' a_2 ' \cdots a_N ' \rangle.
\end{eqnarray}
By definitions the local density operators $\rho_{[i_1 ,i_2 ,\cdots ,i_n]}$
 are non-negative, i.e., $\rho_{[i_1 ,i_2 ,\cdots ,i_n]}\geq 0$. 
For each $\rho_{[i_1 ,i_2 ,\cdots ,i_n]}$,
 we can introduce a partial SLD operator
  ${\cal L}_{[i_1 ,i_2 ,\cdots ,i_n]}$
 as
\begin{equation}
\partial_g \rho_{[i_1 ,i_2 ,\cdots ,i_n]}
=\frac{1}{2} ({\cal L}_{[i_1 ,i_2 ,\cdots ,i_n]}
\rho_{[i_1 ,i_2 ,\cdots ,i_n]}
 +\rho_{[i_1 ,i_2 ,\cdots ,i_n]}
 {\cal L}_{[i_1 ,i_2 ,\cdots ,i_n]} ).
\end{equation}
Then it is possible to rewrite the second information as
\begin{eqnarray}
J^{(N)}&=&\sum_{(i_1 ,i_2 ,\cdots ,i_n)} J^{(N)}_{[i_1 ,i_2 ,\cdots ,i_n]},
\label{300}\\
J^{(N)}_{[i_1 ,i_2 ,\cdots ,i_n]} &=&
Tr\left[
\rho_{[i_1 ,i_2 ,\cdots ,i_n]} {\cal L}^2_{[i_1 ,i_2 ,\cdots ,i_n]}
\right].
\end{eqnarray}
Here $\sum_{(i_1 ,i_2 ,\cdots ,i_n)}$ means the sum over all the subsequences
$(i_1 ,i_2 ,\cdots ,i_n)$ of $(1,2,3,\cdots,N)$, 
including the empty subsequence $\o$.
The decomposition representation makes the evaluation of $J^{(N)}$ 
 easier in many practical applications by using a 
 useful formula for the operators
   $\rho_{[i_1 ,i_2 ,\cdots ,i_n]}$ in the next section.

\section{Evaluation of the Local Density Operator }
\ \\

 The accessible density operators $\rho^{(N)}_\parallel$ 
 can be followed by our apparatus,  
 since the operators  $\rho_\parallel^{(N)}$ 
 are completely local by definition. 
Meanwhile, the local density operator $\rho^{(N)}$ has been so far 
defined 
based upon the total density operator $\rho_{tot}^{(N)}$
 in the previous section. 
 We must say that the definition is   too formal  
 from the  the practical viewpoint,
 because we seldom know global information about the 
  total density $\rho_{tot}^{(N)}$ 
   due to the limitation
 of our ability to measure the system. 
  For the realistic evaluation of $J^{(N)}$, 
  it is convenient to write down $\rho^{(N)}$ explicitly 
  in terms of locally accessible quantities 
  just as the operator $\rho^{(N)}_\parallel$. Such a reformulation can be realized for the cases with factorized evolutions, i.e.,
   $U^{(N)}(t) = [U(t)]^{\otimes N}$ as follows. It
    should be emphasized that we do {\em not}  need to assume 
the i.i.d.  condition for the initial density operator.

Suppose that a composite system $\bS^{\otimes N}$ of N identical $\bS$ 
subsystems is governed by a unitary evolution and that the 
evolution is factorized for each subsystem $\bS$:
\begin{equation}
\rho_{tot}^{(N)} (t) =[U(t)]^{\otimes N} \rho^{(N)}_{tot} (0)
[U(t)^\dagger ]^{\otimes N},
\end{equation}
where $U(t)$ is the unitary time evolution operator for $\bS$ 
and $\rho_{tot}(0)$ is arbitrary initial density operator, which may have
 entanglement between the subsystems.

Let $\cO_{\cM}:=\{e_a |e_a^\dagger =e_a,\ P e_a P =e_a \}$ 
 denote the complete basis  of the available observables acting on $\cM$ 
 for each subsystem $\bS$. 
Even in our local experiments,
 we are able to define and measure the projective evolutions 
 for the available observables $e_a$.
 The  projective
 evolutions are given by stochastic mappings $\Gamma(g,t)[e_a]$ which are 
 defined by
\begin{equation}
\Gamma(g,t)[e_a] :=PU(t) e_a U(t)^\dagger P.
\label{203}
\end{equation}
In various physical systems, 
 the dynamics is  first given  by not $\Gamma(g,t)$ but
  a Lindblad differential form given by
\begin{equation}
\partial_t \rho_\parallel =T_g [\rho_\parallel]
\end{equation}
for arbitrary density operators $\rho_\parallel$
  on $\cM$. 
Here $T_g$ is a time-independent Lindblad  super-operator.
Then the super-operator $T_g$ 
 is related formally to the stochastic mapping $\Gamma(g,t)$
 via
\begin{equation}
T_g = \partial_t \Gamma(g,t=0).
\end{equation}
By integrating formally as $\Gamma(g,t) =e^{tT_g}$, 
 we can recover the stochastic super-operators $\Gamma(g,t)$.

Stress  that 
the operators $\Gamma(g,t)[e_a]$ are completely local quantities we can 
observe. Moreover, 
the projective evolutions for the composite available observables are also
 completely local quantities, which are written as 
\begin{eqnarray}
\Gamma(g,t)^{\otimes k} [e_{a_1}\otimes \cdots \otimes e_{a_k}]
=P^{\otimes k} U(t)^{\otimes k}
(e_{a_1}\otimes \cdots \otimes e_{a_k})
(U(t)^{\otimes k} )^\dagger P^{\otimes k}.
\end{eqnarray}
Our aim in this section is to express 
the operators $ \rho_{[i_1 ,i_2 ,\cdots ,i_n]}$ of the local density operator
 $\rho^{(N)}$  only in terms of 
 the accessible operators like 
$\Gamma(g,t)^{\otimes k} [e_{a_1}\otimes \cdots \otimes e_{a_k}]$.

 Since the initial density operator satisfies $\rho_{tot}^{(N)}(0)
 =P^{(N)} \rho_{tot}^{(N)}(0) P^{(N)}$,  
 the operator $\rho_{tot}^{(N)} (0)$ can be 
 expanded 
 using the basis $\{e_a\}$:
\begin{equation}
\rho_{tot}^{(N)} (0)=\sum_{a_1}\sum_{a_2}\cdots \sum_{a_N}
C_{a_1 a_2 \cdots a_N}
e_{a_1} \otimes e_{a_2} \otimes\cdots \otimes e_{a_N},
\end{equation}
where the real coefficients $C_{a_1 a_2 \cdots a_N}$ is uniquely determined
 by $\rho_{tot}^{(N)} (0)$. 
After rather straightforward calculations, we argue that the following
 relations really hold:
\begin{eqnarray}
\rho_{[i_1 ,i_2 ,\cdots ,i_n]}(t)
&=&
\sum^n_{m=0}(-1)^{n-m}
\sum_{(j_1,\cdots,j_m )\subseteq(i_1 ,\cdots ,i_n) }
\nonumber\\
&&\times
\Tr_{(i_1 ,i_2 ,\cdots ,i_n)-(j_1,\cdots,j_m )}
\left[
\Gamma^{\otimes (N-m) }_{[j_1,\cdots,j_m ]}(g,t)
\left[
\Tr_{(j_1,\cdots,j_m )} [\rho_{tot}^{(N)}(0)]
\right]
\right].\nonumber\\
&&\ \label{100}
\end{eqnarray}
Here 
$\sum_{(j_1,\cdots,j_m )\subseteq(i_1 ,\cdots ,i_n)}$ means
 the sum over all the subsequences $(j_1,\cdots,j_m )$
  of the sequence $(i_1 ,\cdots ,i_n)$.
$Tr_{(j_1,\cdots,j_m )}$ is a trace operation
in terms of the $(j_1,\cdots,j_m )$ degrees of freedom. 
$Tr_{(i_1 ,i_2 ,\cdots ,i_n)-(j_1,\cdots,j_m )}$ means 
a trace operation in terms of the complementary subsequence
to the subsequence $(j_1,\cdots,j_m )$ of $(i_1 ,\cdots ,i_n)$.
When $m=0$, $Tr_{(j_1,\cdots,j_m )}$ is reduced into the identical operation.
The operation $\Gamma^{\otimes (N-m) }_{[j_1,\cdots,j_m ]}$ is the 
time-evolution
 operator for all degrees of freedom removing the 
 $(j_1,\cdots,j_m )$ part. Here it is better to note again that 
 even though the formula include subtractions, 
 all the operators $\rho_{[i_1 ,i_2 ,\cdots ,i_n]}$ are
  non-negative $\rho_{[i_1 ,i_2 ,\cdots ,i_n]}\geq 0$.

In eqn(\ref{100}), note that the operators $\rho_{[i_1 ,i_2 ,\cdots ,i_n]}(t)$
 cannot be evaluated only from the knowledge about the accessible density operator $\rho^{(N)}_\parallel (t)$. It is required to solve  independently 
 time evolutions of many descendant operators,  
$$\Gamma^{\otimes (N-m) }_{[j_1,\cdots,j_m ]}(g,t)
\left[
\Tr_{(j_1,\cdots,j_m )} [\rho_{tot}^{(N)}(0)]
\right].$$

Here let us just draw the outline of the proof using the $\rho_{[1]}$ case.
Substituting eqns(\ref{90}) and (\ref{91}), 
the following manipulation is possible:
\begin{eqnarray}
&&
\langle a_2 a_3 \cdots a_N |
\rho_{[1]}
|a_2 ' a_3 ' \cdots a_N ' \rangle
\nonumber\\
&=&
\langle B a_2 a_3 \cdots a_N |
\rho^{(N)}
|B a_2 ' a_3 ' \cdots a_N ' \rangle 
\nonumber\\
&=&
\langle a_2 a_3 \cdots a_N |
\Tr_1 \left[
({\bf 1}-P)\otimes P^{\otimes (N-1)}
\rho_{tot}^{(N)} 
({\bf 1}-P)\otimes P^{\otimes (N-1)}
\right]
|a_2 ' a_3 ' \cdots a_N ' \rangle.
\nonumber
\end{eqnarray}
Moreover we can rewrite the equation as follows:
\begin{eqnarray}
\rho_{[1]}
=
\Tr_1
\left[
{\bf 1}\otimes P^{\otimes (N-1)}
\rho_{tot}^{(N)} (t)
{\bf 1}\otimes P^{\otimes (N-1)}
\right]
-
\Tr_1\left[
\Gamma^{\otimes N} (g,t)[\rho_{tot}^{(N)} (0)]
\right].\nonumber
\end{eqnarray}
It is noticed that the first term of the r.h.s. is reduced 
using the expansion of $\rho_{tot}^{(N)} (0)$ as 
follows. 
\begin{eqnarray}
&&
\Tr_1
\left[
{\bf 1}\otimes P^{\otimes (N-1)}
\rho_{tot}^{(N)} (t)
{\bf 1}\otimes P^{\otimes (N-1)}
\right]
\nonumber\\
&=&
\sum_{a_1\cdots a_N}
C_{a_1 a_2 \cdots a_N}
\Tr_1 [U(t)e_{a_1} U^\dagger (t)]
(PU(t)e_{a_2}U^\dagger(t) P)
\otimes \cdots \otimes
(PU(t)e_{a_N}U^\dagger (t)P)
\nonumber\\
&=&
\Gamma^{\otimes {(N-1)}}(g,t)\left[
\sum_{a_1\cdots a_N}
C_{a_1 a_2 \cdots a_N}
\Tr_1 [e_{a_1}]e_{a_2}\otimes \cdots 
\otimes e_{a_N} \right]
\nonumber\\
&=&
\Gamma^{\otimes {(N-1)}}(g,t)[\Tr_1[\rho^{(N)}_{tot} (0)]].
\end{eqnarray}
Consequently we arrive at the relation for $\rho_{[1]}$
 in eqn(\ref{100}):
\begin{equation}
\rho_{[1]} =
\Gamma^{\otimes {(N-1)}}(g,t)[\Tr_1[\rho^{(N)}_{tot}(0)]]
-
\Tr_1\left[
\Gamma^{\otimes N} (g,t)[\rho_{tot}^{(N)} (0)]
\right].
\end{equation}
The proofs for the other components in eqn(\ref{100}) can be achieved
 in the similar ways.

The relation in eqn(\ref{100}) makes the evaluation of $J^{(N)}$ possible,
 only based upon our local knowledge.

\section{ A Decaying Two-Level Model with a Small Unknown Parameter}
\ \\

In order to demonstrate our formulation explicitly, 
 let us consider a  system including 
 a small unknown parameter $g$. In many physical systems, 
 the estimation of such  a small parameter often 
 provides significant physical information. For example, 
 tiny coupling constants in the elementary particle interactions 
 produce only quite rare processes, however, the analyses give a lot of important constraints of high energy
 features beyond the today's accelerator technology. 
 For simplicity suppose a decaying two-level system including the small 
 coupling $g$.  The model has been frequently 
  used in physics, for instance, to analyze the flavor-oscillating
  phenomena in the $K_0 -\bar{K}_0$ system \cite{P}.
The Hamiltonian of the example is given as 
\begin{equation}
H=-i\hbar
\left[
\begin{array}{cc}
\Gamma_+ & 0 \\
0 & \Gamma_-
\end{array}
\right]
+g\hbar
\left[
\begin{array}{cc}
0 & 1 \\
1 & 0
\end{array}
\right],
\end{equation}
where $\Gamma_\pm >0 $, $\Gamma_+ \neq \Gamma_-$ and 
$|g| \ll \Gamma_\pm, \ |\Gamma_+ -\Gamma_-| $.
In the two-level 
subspace, time evolution of the density matrix $\rho_\parallel$
 is governed by the following equation of motion:
\begin{equation}
i\hbar \partial_t \rho_\parallel 
=H\rho_\parallel -\rho_\parallel H^\dagger.\label{304}
\end{equation}
 
Define the states $|\pm\rangle$ as
\begin{equation}
\sigma_z |\pm\rangle =\pm |\pm\rangle.
\end{equation}
Here $\sigma_z$ is the z-component of the Pauli matrix. 
Let us estimate the time evolution of $|\pm\rangle$ in the first order of $g$.
It is solved as
\begin{eqnarray}
|\pm (g,t) \rangle
=
e^{-\Gamma_\pm t} |\pm\rangle
+ig d(t)|\mp\rangle
 +O(g^2) ,
\end{eqnarray}
where the functions $d(t)$ is given as
\begin{equation}
d(t)=\frac{e^{-\Gamma_+ t} -e^{-\Gamma_- t}}{\Gamma_+ -\Gamma_-}.
\end{equation}
Then the information $J_\pm$ for the state $|\pm (t)\rangle$ 
is evaluated as 
\begin{eqnarray}
J_\pm(t) =
4d(t)^2+O(g). \label{301}
\end{eqnarray}
The Cram\'er-Rao bound is always achieved by an observable
\begin{equation}
A(t_\ast ) =\sigma_y +O(g),
\end{equation}
where $\sigma_y$ is the y-component of the Pauli matrix.

In this simple model we are able to optimize the measurement time $t$.
The Fisher information takes its maximum value in the lowest order 
\begin{equation}
J_{max}
=\frac{4}{(\Gamma_+ -\Gamma_- )^2}
\left(
\left(\frac{\Gamma_-}{\Gamma_+}\right)^{\frac{\Gamma_+}{\Gamma_+ -\Gamma_-}}
-
\left(\frac{\Gamma_+}{\Gamma_-}\right)^{\frac{\Gamma_-}{\Gamma_- -\Gamma_+}}
\right)^2 +O(g),
\end{equation}
at
\begin{equation}
t_\ast= \frac{\ln \Gamma_+ -\ln \Gamma_-}{\Gamma_+ -\Gamma_-}+O(g).\label{50}
\end{equation}

In eqn(\ref{50}), when $\Gamma_-$ is much smaller than  $\Gamma_+$:
\begin{equation}
\Gamma_- \ll \Gamma_+,
\end{equation}
$t_\ast \sim -\frac{1}{\Gamma_+}\ln\frac{\Gamma_-}{\Gamma_+} $
becomes larger logarithmically. 
In fact,
 the time $t_\ast$ can be late until the first order estimation breaks down,
the time $t_g \sim -\frac{1}{\Gamma_+}\ln\frac{g}{\Gamma_+}$.
Interestingly, at $t=t_\ast$,
 the survival probability for the state $|+(g, t)\rangle$ 
 in the two-level subspace is estimated as
\begin{equation}
\langle +(g, t_\ast) |+ (g, t_\ast) \rangle \sim
\left(\frac{\Gamma_-}{\Gamma_+} \right)^2 \ll 1.
\end{equation}
Against a naive expectation, this indicates 
that the best quantum estimation can be achieved at the
 time after the state has almost escaped from the two-level subspace.

Analysis of a composite system of the two subsystems may be also instructive.
Let us first take the initial state as an i.i.d. state,
\begin{equation}
|\Psi (0) \rangle = |++\rangle.
\end{equation}
For the density matrix $\rho^{(2)}_\parallel (t)=
|++(t)\rangle\langle ++ (t) |$, the fisher information $j^{(2)}$
 defined by

\begin{equation}
j^{(2)}:
=\Tr[ \rho_{\parallel}^{(2)} (L^{(2)})^2]
+
\frac{ [ \Tr [ \rho_{\parallel}^{(2)} L^{(2)} ] ]^2}
{1-\Tr\rho_{\parallel}^{(2)} }
\end{equation}

is evaluated as
\begin{equation}
j^{(2)} (t) =8d(t)^2 e^{-2\Gamma_+ t} +O(g).
\end{equation}
Because $j^{(1)} =4d^2 +O(g)$, the relation $j^{(2)}(t) =2j^{(1)}$ is not satisfied due to the $e^{-2\Gamma_+ t}$ factor and $j^{(2)}$ is exponentially 
 smaller than $2j^{(1)}$ at $t>0$.

Let us compare the result with $J^{(2)}$ defined by eqn(\ref{300}).
In this case each  component 
of the local density operator is defined  as
\begin{eqnarray}
\rho_{[\o]}(t)&=& \Gamma^{\otimes 2}(g,t)[\rho(0)]=\rho_\parallel^{(2)} (t),\\
\rho_{[1]}(t)&=&
\Gamma(g,t)[\Tr_1[\rho(0)]]
-
\Tr_1\left[
\Gamma^{\otimes 2} (g,t)[\rho(0)]
\right],\\
\rho_{[2]}(t)&=&
\Gamma(g,t)[\Tr_2 [\rho(0)]]
-
\Tr_2\left[
\Gamma^{\otimes 2} (g,t)[\rho(0)]
\right],\\
\rho_{[1,2]}(t)&=&1-
\Tr_2\left[
\Gamma(g,t)[\Tr_1[\rho(0)]]
\right]
-
\Tr_1\left[
\Gamma(g,t)[\Tr_2[\rho(0)]]
\right]
\nonumber\\
&&
+
\Tr_{1,2}
\left[
\Gamma^{\otimes 2} (g,t)[\rho(0)]
\right].
\end{eqnarray}
As seen above, 
to calculate $\rho_{[1]}$, $\rho_{[2]}$ and $\rho_{[1,2]}$, 
we need the
time evolution of the partial density matrices 
$\Gamma(g,t)[\Tr_1[\rho(0)]]$ and $\Gamma(g,t)[\Tr_2[\rho(0)]]$. 
It should be stressed that these evolutions cannot 
be obtained only from knowledge of $\rho_{\parallel}^{(2)}(t)$, for instance,
 by taking any traces for$\rho^{(2)}_\parallel$. 
 They must be calculated independently 
by solving eqn(\ref{304}) for the
 initial density matrices $\Tr_1[\rho(0)]$ 
 and $\Tr_2[\rho(0)]$.  
For the initial i.i.d. density matrix, 
each $J^{(2)}$ component is calculated as
\begin{eqnarray}
J^{(2)}_{[\o]} &=&j^{(2)}=8d(t)^2 e^{-2\Gamma_+ t} +O(g),
\\
J^{(2)}_{[1]} &=& 4d(t)^2 (1-e^{-2\Gamma_+ t}) +O(g),
\\
J^{(2)}_{[2]} &=& 4d(t)^2 (1-e^{-2\Gamma_+ t}) +O(g),
\\
J^{(2)}_{[1,2]} &=& O(g).
\end{eqnarray}
Thus total information $J^{(2)}$ is precisely equal to twice of $J^{(1)}$:
\begin{equation}
J^{(2)} =8d(t)^2 +O(g) =2J^{(1)}.
\end{equation}

Next let us discuss an entangled case.
Initially we take a state as
\begin{equation}
|\Phi (0) \rangle =
\frac{1}{\sqrt{2}}
[|+-\rangle + |-+\rangle].
\end{equation}
Calculation of $j^{(2)}(t)$ is easy and the results are as follows.
\begin{eqnarray}
j^{(2)} (t)
=
8d(t)^2 [e^{-2\Gamma_+ t} +e^{-2\Gamma_- t}] +O(g).
\end{eqnarray}
Note that $j^{(2)} (t)/j^{(1)} (t)$ vanishes exponentially in time 
 just as in the i.i.d. case.

Evaluation of $J^{(2)} (t)$ needs not only  
the density matrix
\begin{equation}
\rho_{\parallel } (t) =|\Phi (t) \rangle\langle \Phi (t)|
\end{equation}
 but also another density matrix
\begin{eqnarray}
\Gamma(g,t)[\Tr_1[\rho(0)]]=\Gamma(g,t)[\Tr_2[\rho(0)]]
=\Gamma(g,t)\left[\frac{1}{2}{\bf 1} \right].
\end{eqnarray}
After some manipulations  the form  of $J^{(2)} (t)$ results in 
\begin{eqnarray}
&&
J^{(2)} (t)
\nonumber\\
&=&
8d(t)^2 [e^{-2\Gamma_+ t} +e^{-2\Gamma_- t}]
\nonumber\\
&&
+4d(t)^2
[1+2e^{-(\Gamma_+ +\Gamma_-)t}]^2
\frac{[e^{-\Gamma_+ t} -e^{-\Gamma_- t}]^2}
{e^{-2\Gamma_+ t} (1-e^{-2\Gamma_- t})
+
e^{-2\Gamma_- t} (1-e^{-2\Gamma_+ t})
}
\nonumber\\
&&+O(g).
\end{eqnarray}

Note that at the early era ($t\sim 0$), both 
$j^{(2)} (t)$ and $J^{(2)} (t)$ 
have
four-times information compared with the single system:
\begin{eqnarray}
&&
j^{(2)} (t\sim 0 )\sim 4j^{(1)},
\\
&&
J^{(2)} (t\sim 0 )\sim 4J^{(1)} . 
\end{eqnarray}
Thus the information is twice larger than the above i.i.d. case. 
 Obviously this advantage arises due to the entanglement between subsystems.

For the entangled case, $J^{(2)} /J^{(1)}$  becomes smaller 
than the value of the i.i.d. case (equal to two ) in the late time.
Hence, the i.i.d. density operator 
becomes more relevant than  the entangled
 density operator for the estimation of $g$.
In the limit of $t\rightarrow \infty$, 
the value of $J^{(2)} /J^{(1)}$ for the entangled case 
  approaches to  
the single-system value: 
\begin{equation}
\lim_{t\rightarrow\infty}
\frac{J^{(2)} (t)}{J^{(1)} (t)} \sim  1 .
\end{equation}
This is due to contributions of the one-blank states
 ($|B\pm\rangle$ and $|\pm B\rangle$). 
Consequently, it can be said that 
 the measurement should be at the early times
 in order to utilize  
 enhancement of the Fisher information by the entanglement.

So far we have discussed only systems with  small numbers of samples. 
For the practical estimation of the small parameter $g$, 
the many-samples estimation is
 inevitable beyond the above simple examples. 
For instance, the minimized expected error $\delta g$  
is given by 
\begin{equation}
\delta g=\frac{1}{\sqrt{NJ^{(1)}}}
\end{equation}
for the i.i.d. cases of $N$-samples systems. 
Then, in order to get a meaningful estimate,
 the number of the samples must be,  at least, 
 $O\left(1/(g^2 J^{(1)}) \right)$ for the correct value $g$. 
 It is expected that large entanglement between  
 many samples may  extremely improve the estimation for $g$  
 and make the  number of the samples enough for the estimation much smaller. 

\ \\
\ \\
\ \\

\section{Summary}
\ \\

We have investigated
 deeply the local quantum estimation problem of an unknown parameter.
 The practical restriction of  experimental observables 
 takes place in various situations of the physical experiments.
For a typical example, 
in particle physics we can probe only low-energy visible sectors
 of the whole system by our present devices. Such obstacles 
 appear  because of the limit of the present technology
  and so on.  Moreover,
   observation of quantum phenomena, which  
  happen only at  quite small rates,  
  often 
 becomes  the crucial target of experiments, which may derive
  some profound results of physics like, for instance, CPT violation \cite{HP}.
 In such situations, 
 it is generally difficult to take a large number of data as one wishes,
  at least, in the first stage of the experimental studies.
Hence the local quantum estimation becomes really important when  
 the experimental arrangements are designed, because the estimation theory
  provides among our available probes the  optimized observable  
  which quantum fluctuation is most suppressed 
  in the estimation based upon
 a limited number of the data.

In spite of such relevance of the local quantum estimation,
 the problem 
 has never been discussed in detail, as far as the authors know.
In this paper, the detailed analysis and formulations based upon
 the Fisher informations have been completed. 
After a brief review on the standard quantum estimation theory,
 the local quantum estimator for the local estimation has been 
  defined
 by eqn(\ref{40}).  
 The notion of the local density operators  
 was clearly introduced in eqn(\ref{e20}),
 and the Cram{\'e}r-Rao inequality in the local quantum estimate theory
 (eqn(\ref{cr})) has been proven by taking  
 the local Fisher information defined by eqn(\ref{34}).
 The inequality is a fundamental tool in the theory and 
  will play a significant role in the local estimation 
 in various physical applications. 
  In section 6, the Fisher 
 information for the unnormalized pure state was commented. 
 The formula in eqn(\ref{unps}) is an extension of that derived by
 Fujiwara and Nagaoka, who discussed the Fisher information for
  normalized pure states. It is known that in many physical systems  
  non-unitary theories of pure states also are available and
 that  the validity is well verified by the experiments.
 In such systems with non-unitary evolution,
 eqn(\ref{unps}) is quite useful to
  evaluate the Fisher information for an unknown parameter.
 In  section 7, it was pointed out that 
 the local quantum estimation in the composite system has two
  independent formulations, using the i.i.d.cases. 
  In  section 8, two general formulations of 
 the local quantum estimation for the composite system were proposed. 
For the composite system of $N$ identical subsystems, 
we have two Fisher informations, $j^{(N)}$ and $J^{(N)}$. 
 The information $j^{(N)}$ 
 takes a simple form to define, 
  but  gives, in general,  much smaller values than $J^{(N)}$. 
 The theory of information $J^{(N)}$  
 can generate a more precise estimate for $g$, but has a pretty complicated 
 form to deal with, compared to the $j^{(N)}$ case. 
 In order to avoid the troublesome procedures in evaluation of $J^{(N)}$, 
  we showed in  section 9 the formula in eqn(\ref{100}), 
  which makes the evaluation  tractable.  
   As seen in eqn(\ref{100}),
 calculation of the Fisher information $J^{(N)}$ requires 
 solving evolutions of many descendant operators, 
 $\Gamma^{\otimes (N-m) }_{[j_1,\cdots,j_m ]}(g,t)
\left[
\Tr_{(j_1,\cdots,j_m )} [\rho_{tot}^{(N)}(0)]
\right]$, independently of solving the accessible density operator
 $\rho^{(N)}_\parallel (t)$ itself. Such 
 processes never appear in the ordinary quantum estimation theory,
 where the Fisher information can be evaluated  by using only 
 a time-evolved density operator. 
 In  section 10, we demonstrated explicitly our formulation 
 of the local quantum estimation by applying 
 to a decaying two-level system with a small unknown parameter.

 We hope that 
 the analysis in this  paper enables the quantum estimation theory to
 take a more active part in the real experimental studies, which 
 suffer from the restriction of available observables and the practical 
 limitation of the number of the data.

\section*{Acknowledgements}

The authors thank M.Hayashi for helpful comments. 
One of the authors (M.O.) is supported by the Strategic Information and Communications 
R\&D Promotion Scheme of the MPHPT of Japan, by the CREST project of the JST, and by the
Grant-in-Aid for Scientific Research of the JSPS.

\section*{Appendix}

In this appendix, the Cram{\'e}r-Rao inequality is proved. 
Let us write the triangular inequality relation as
\begin{equation}
\Tr (X^\dagger X)\Tr(Y^\dagger Y) \geq |\Tr(X^\dagger Y)|^2,\label{e6}
\end{equation}
where $X$ and $Y$ are arbitrary operators acting on the Hilbert space.
Decomposing the operator $X^\dagger Y$ into the sum 
of the real and imaginary parts 
as
\begin{equation}
X^\dagger Y = \frac{1}{2} (X^\dagger Y +Y^\dagger X)
+\frac{1}{2}(X^\dagger Y -Y^\dagger X),
\end{equation}
another inequality relation arises:
\begin{eqnarray}
&&
\Tr (X^\dagger X)\Tr(Y^\dagger Y) \geq 
\frac{1}{4}|\Tr(X^\dagger Y +Y^\dagger X)|^2
+
\frac{1}{4}|\Tr(X^\dagger Y -Y^\dagger X)|^2
\nonumber\\
&&
\geq \frac{1}{4}|\Tr(X^\dagger Y +Y^\dagger X)|^2.\label{e7}
\end{eqnarray}
Here let us take
\begin{eqnarray}
&&
X=L(g)\sqrt{\rho_{tot} (g)},\label{e13}
\\
&&
Y=(A -E_g [A] )\sqrt{\rho_{tot} (g)}.\label{e14}
\end{eqnarray}
Then, from the inequality (\ref{e7}), we can derive that
\begin{eqnarray}
&&
\Tr[\rho_{tot} (g)L(g)^2]
\Tr[\rho_{tot} (g) (A -E_g [A] )^2]
\nonumber\\
&&
\geq
\frac{1}{4}
\left|
\Tr
\left[\rho_{tot} (g) \left(L(g) (A-E_g[A]) +(A-E_g [A] ) L(g) \right) \right]
\right|^2.\label{e10}
\end{eqnarray}
The right-hand-side term in the above inequality is able to be calculated
 using eqns(\ref{2e}), (\ref{eq:SLD}) and (\ref{1e}) successively as follows.
\begin{eqnarray}
&&
\frac{1}{4}
\left|\Tr
\left[\rho_{tot} (g) \left(L(g) (A-E_g[A]) +(A-E_g [A] ) L(g) \right) \right]
\right|^2
\nonumber\\
&=&
\frac{1}{4}
\left|\Tr
\left[\rho_{tot} (g) \left(L(g) A +A L(g) \right) \right]
\right|^2
\nonumber\\
&=&
\frac{1}{4}
\left|\Tr
\left[ A\left(\rho_{tot} (g) L(g)  + L(g) \rho_{tot} (g)\right) \right]
\right|^2
\nonumber\\
&=&
\left(
\Tr[A \partial_g \rho_{tot} (g) ]
\right)^2 = \left( \partial_g E_g [A] \right)^2.
\end{eqnarray}
Consequently the relation (\ref{e10}) 
 implies the following inequality:
\begin{equation}
 \frac{V_g [A]}{\left( \partial_g E_g [A] \right)^2} \geq \frac{1}{J_g},
 \label{e15}
\end{equation}
 thus, the inequality  (\ref{e111}) is proved. For the unbiased case with
 $E_g [A] =g$, the inequality(\ref{e15}) is reduced to 
  (\ref{crie}). The equality is trivially
 attained when $X\propto Y$ in eqn(\ref{e7}) and the relation $X\propto Y$ 
  holds in eqns (\ref{e13}) and (\ref{e14}) 
  when  we set  $A\propto L(g)$, because  $E_g [L(g)]=0$.

\end{document}